\documentclass[twocolumn,prb,amsmath,amssymb,showpacs,
superscriptaddress,floatfix]{revtex4}
\usepackage{hyperref}
\usepackage{graphicx}
\usepackage {amsmath}
\DeclareMathOperator{\Tr}{Tr} \DeclareMathOperator{\C60}{C_{60}} 
\begin{document}

\title{Orientational glass transition in $\C60$.}

\author{T.I. Schelkacheva}
\affiliation{Institute for High Pressure Physics, Russian Academy of Sciences, Troitsk 142190,
Moscow Region, Russia}

\author{E.E. Tareyeva}
\affiliation{Institute for High Pressure Physics, Russian Academy of Sciences, Troitsk 142190,
Moscow Region, Russia}

\author{N.M. Chtchelkatchev}
\affiliation{Institute for High Pressure Physics, Russian Academy of Sciences, Troitsk 142190,
Moscow Region, Russia}

\date{\today}

\begin{abstract}
We construct a model for qualitative description of the orientational glass transition in $\C60$
on the spin-glass theory basis. The physical origin of the frustration and the disorder is
discussed.
\end{abstract}


\maketitle

As is established in a number of experiments, $\C60$ crystalizes in a face centered cubic (fcc)
structure. At ambient temperature the molecules rotate almost freely with centers on the fcc
lattice sites, so that the space group is $Fm\overline{3}m$. When the temperature decreases to
$T_c\approx 260\, \mathrm{K}$ the first order orientational phase transition takes place: the
sites of the initial fcc lattice become divided between four simple cubic sublattices  with its
own preferable molecular orientation in each sublattice. The broken symmetry group is
$Pa\overline{3}$.

Moreover the neutron-diffraction experiments [see, e.g., reviews \onlinecite{Sundqvist,Moret}]
have shown that the orientations in the ordered state are so that the electron-rich regions (the
interpentagon double bonds) face the electron-deficient regions of the the neighboring $\C60$
molecule: the centers of pentagons or the centers of hexagons.

Solid $\C60$ undergoes a well-known glass transition at $T_g=90$K
when no orientational motion can usually be
detected.\cite{Sundqvist,Moret,Gugenberger,David3,David,David2,Aksenova,Aleksandrovskii,Lebedev}
Each $\C60$ molecule has a single misorientation separated from
the ground-state orientation by a large energy barrier $E_g$. The
energy difference between these two orientations is $U$ and
$E_g\gg U$.\cite{Sundqvist,Gugenberger,David3,Yildirim,Yu} These
two minima of the intermolecular angle dependent energy were shown
to be much lower than the energies of other mutual orientations of
the pair of molecules.\cite{Moret,Chaplot,Haunois} Below $T_g$
molecules orientations are forming glass and the occupation
probability of misoriented molecules is essentially frozen at
value of $\approx 17\%$. And this static orientational disorder
persists down to very low temperatures. The thermal energy is too
small compared with the energy threshold  between the two states
for further reorientation to be possible.

In spite of the recent progress the present understanding of the intermolecular interactions in
solid $\C60$ is still imperfect and no single model is able to describe correctly its whole
physical properties.\cite{Chaplot,Haunois} This paper discusses glassy orientational transition in
$\C60$. The glass state in $\C60$ usually was described purely phenomenologically as ``freezing''
of an ensemble of two-level systems. The step towards microscopic description of glassy
orientational transitions in ortho-para hydrogen was developed recently in
Refs.\onlinecite{Luchinskaya,Schelkacheva}; the developed approach is useful because it takes
carefully into account disorder and frustration. The symmetry of $\C60$ molecules essentially
differs from the symmetry of hydrogen; the method used in
Refs.\onlinecite{Luchinskaya,Schelkacheva} can not be directly used for glassy orientational
transitions in $\C60$. We show how the approach of Refs.\onlinecite{Luchinskaya,Schelkacheva}
should be modified for $\C60$ case.

Now we follow the consideration of the Ref.\onlinecite{Tareyeva1}.
We consider the restricted number of allowed orientations instead
of the continuous rotations. Let us take into account in the
energy only the orientations with pentagons, hexagons or double
bonds directed towards 12 nearest neighbors in fcc-lattice. The
$\C60$ molecule is constructed in such a way that if 6 of its 12
pentagons (or 6 of its 20 hexagons) face 6 nearest neighbor double
bonds (P and H states) then 6 of its 30 interpentagon double bonds
face the remaining nearest neighbors. Now the energy matrix
elements can take only three values; $J_0$, the energy of the
general mutual position, $J_p$ pentagon versus double bond, and
$J_H$, hexagon versus double bond. Following
Ref.\onlinecite{Lapinskas} and putting $J_0=0$ we obtain from
Fig.(2b) of Ref.\onlinecite{Yildirim} $J_P=-300$K and $J_H=-110$K.
The large number of the nearest neighbor bonds allows for the
mean-field description of the problem.

We do not use the multipole expansion. So in the framework of our model calculations it is
possible to build up the allowed functions using only the harmonics with $l=6$. We need only their
transformation properties and effectively take into account higher order terms.

Let us construct the functions $P_i(\omega)$ and $H_i(\omega)$
explicitly in terms of cubic harmonics $K_m=K_{6,m}$,
$m=1,2,\ldots,13$ (see, e.g., Ref.\onlinecite{Heid}). All
functions $P_i$ and $H_i$ are the sums of $K_m$, invariant under
the icosahedral symmetry of the molecule (i.e., belonging to the
$A_{1g}$ representation of the icosahedral group $I_h$) if
icosahedrons are naturally oriented in one of 8 properly chosen
coordinate systems. The states $P_i(H_i)$ have 6 pentagons
(hexagons) and 6 double bonds directed towards 12 nearest
neighbors along different $[100]$ axes. $P_1(\omega)$ describes
the molecule rotated from the standard orientation $B$ (following
Ref.\onlinecite{Harris}) about $[111]$ axis through the angle
$97.76125^\circ$. The angle for $H_1(\omega)$ is $37.76125^\circ$.
The functions $P_2(\omega)$, $P_3(\omega)$ and $P_4(\omega)$ (or
$H_2$, $H_3$, $H_4$) are obtained from $P_1(\omega)$ ($H_1$) by
subsequent counterclockwise rotations of the molecule by
$90^\circ$ around ${z}$ axes.

If written in standard coordinate frame with cartesian axes along the cube sites these functions
have the following explicit form:
\begin{multline}\label{eq:P1}
P_1(\omega)=\alpha_P K_1(\omega)+\beta_P[K_8(\omega)+K_9(\omega)+K_{10}(\omega)]
\\
+\gamma_P[K_{11}(\omega)+K_{12}(\omega)+K_{13}(\omega)],
\end{multline}
\begin{multline}\label{eq:P2}
P_2(\omega)=\alpha_P K_1(\omega)+\beta_P[-K_8(\omega)+K_9(\omega)-K_{10}(\omega)]
\\
+\gamma_P[-K_{11}(\omega)+K_{12}(\omega)-K_{13}(\omega)],
\end{multline}
\begin{multline}\label{eq:P3}
P_3(\omega)=\alpha_P K_1(\omega)+\beta_P[K_8(\omega)-K_9(\omega)-K_{10}(\omega)]
\\
+\gamma_P[K_{11}(\omega)-K_{12}(\omega)-K_{13}(\omega)],
\end{multline}
\begin{multline}\label{eq:P4}
P_4(\omega)=\alpha_P K_1(\omega)+\beta_P[-K_8(\omega)-K_9(\omega)+K_{10}(\omega)]
\\
+\gamma_P[-K_{11}(\omega)-K_{12}(\omega)+K_{13}(\omega)],
\end{multline}
with $\alpha_P=-0.38866$; $\beta_P=0.31486$; $\gamma_P=-0.42877$. The functions $H_i(\omega)$ have
the same form as $P_i(\omega)$ but with the coefficients $\alpha_H=0.46588$; $\beta_H=0.37740$;
$\gamma_H=0.34432$. The functions are normalized to unity.

In the mean-field approximation by minimizing the orientational free energy one can obtain the
nonlinear integral equation for the one-particle orientational distribution functions
$g_i(\omega)$ for a molecule on ith sublattice of fcc $\C60$. In the neighborhood of the
bifurcation point $T_b$ we have a linearized system:
\begin{multline}\label{eq:h1}
h_1(\omega)+\frac 1{4\pi T}\int d\omega'\left[B(\omega,\omega') h_2(\omega')+A(\omega,\omega')
h_3(\omega')\right.
\\
\left.+D(\omega,\omega') h_4(\omega')\right]=0,
\end{multline}
where $g_i(\omega)=\mu h_i(\omega)+\ldots$; $\mu=1/T-1/T_b$ and analogously for three other
sublattices.

Here $A(\omega,\omega')$, $B(\omega,\omega')$, $D(\omega,\omega')$ are the sums of interactions
over the nearest neighbors in the sublattices. For example, the sum in the plane perpendicular to
the $x$ axis can be written explicitly in the form
\begin{gather}\label{eq:D}
\begin{split}
D(\omega,\omega'&)=2\left\{\left[\left(P_1(\omega)+P_4(\omega)\right)J_P+
(H_1(\omega)+H_4(\omega))J_H\right]\right.
\\
&\left.\times\left[P_2(\omega')+P_3(\omega')+H_2(\omega')+H_3(\omega')\right]\right.
\\
+&\left.\left[P_2(\omega)+P_3(\omega)+H_2(\omega)+H_3(\omega)\right]\right.
\\
&\left.\times\left[(P_1(\omega')+ P_4(\omega'))J_P+(H_1(\omega')+H_4(\omega'))J_H\right]\right.
\\
+&\left.\left[(P_2(\omega)+P_3(\omega))J_P+(H_2(\omega)+H_3(\omega))J_H\right]\right.
\\
&\left.\times\left[P_1(\omega')+ P_4(\omega')+H_1(\omega')+H_4(\omega')\right]\right\}
\\
+&\left.\left[P_1(\omega)+P_4(\omega)+H_1(\omega)+H_4(\omega)\right]\right.
\\
&\left.\times\left[(P_2(\omega')+ P_3(\omega'))J_P+(H_2(\omega')+H_3(\omega'))J_H\right]\right\}.
\end{split}
\end{gather}
We add the condition $h_i(\omega)$ to transform one into another under the action of the cubic
group rotation elements which leave the fcc lattice invariant. At the bifurcation point $T_b>0$
nontrivial solution with broken symmetry appears, corresponding to the orientationally ordered
phase. We have $T_b=275$K ($T_c^{(\mathrm{exp})}\approx260$K) and
\begin{gather}\label{eq:h}
h_1(\omega)=a P_1(\omega)+bH_1(\omega)+c K_1(\omega),
\\
h_2(\omega)=a P_3(\omega)+bH_3(\omega)+c K_1(\omega),
\\
h_3(\omega)=a P_4(\omega)+bH_4(\omega)+c K_1(\omega),
\\\label{eq:h4}
h_4(\omega)=a P_2(\omega)+bH_2(\omega)+c K_1(\omega),
\end{gather}
\begin{gather}\label{eq:abQ}
a\alpha_P+b\alpha_H+c=0.
\end{gather}

So we obtain\cite{Tareyeva1} the bifurcation temperature, the
symmetry of the ordered phase and the ratio $\rho$ of the number
of molecules with pentagon facing neighbor double bond near the
phase transition in good agrement with the experimental data:
\begin{gather}
\rho_P=\frac{a}{a+b}=0.608,
\\
\rho_H=\frac{b}{a+b}=0.392.
\end{gather}

The experiments show [see, e.g., Refs.\onlinecite{Sundqvist,Moret}] that the ratio of the number
of molecules in those two states is about $60:40$ at the phase transition temperature, and
increases when the temperature decreases and freezes at $T_g$.

It is convenient to reformulate our results. In the orientational ordered phase all sublattices
are physically equivalent: they convert to each other when we rotate the crystal on $90^\circ$
along the $z$-axis. The expressions for $h_2$, $h_3$ and $h_4$ formally coincide with
$h_1(\tilde\omega)$ in their ``own'' system of coordinates that we label by tilde.  Then we can
write Eqs.\eqref{eq:h1} as a single equation as follows:
\begin{gather}\label{eq:h_tilde_omega}
h_1(\tilde\omega)+\frac1{4\pi T}\int
d\tilde\omega'E(\tilde\omega,\tilde\omega')h_1(\tilde\omega')=0,
\end{gather}
where
\begin{multline}
\int d\omega'\left[B(\omega,\omega')h_2(\omega')+A(\omega,\omega')h_3(\omega')+\right.
\\
\left.+D(\omega,\omega')h_4(\omega')\right]\equiv\int d\tilde\omega'
E(\tilde\omega,\tilde\omega')h_1(\tilde\omega').
\end{multline}

The matrix elements of $E$ are symmetrical. They depend on $J_p$, $J_H$ and the coefficients
$\alpha_{P(H)}$, $\beta_{P(H)}$, $\gamma_{P(H)}$. The E-matrix can be diagonalized:
\begin{gather}\label{eq:tilde_E}
\begin{split}
E(\tilde\omega,\tilde\omega')=\frac 1 {4\pi
T_b}\biggl\{E_r&\,[K_8(\tilde\omega)+K_9(\tilde\omega)+K_{10}(\tilde\omega)]+
\\
+&E_S[K_{11}(\tilde\omega)+K_{12}(\tilde\omega)+K_{13}(\tilde\omega)]\biggr\}\times
\\
\biggl\{E_r&\,[K_8(\tilde\omega')+K_9(\tilde\omega')+K_{10}(\tilde\omega')]+
\\
+&E_S[K_{11}(\tilde\omega')+K_{12}(\tilde\omega')+K_{13}(\tilde\omega')]\biggr\}.
\end{split}
\end{gather}

The solution of the Eq.\eqref{eq:h_tilde_omega} is the following:
\begin{multline}\label{eq:h_solution}
h(\tilde\omega)=r\,[K_8(\tilde\omega)+K_9(\tilde\omega)+K_{10}(\tilde\omega)]+
\\
+s\,[K_{11}(\tilde\omega)+K_{12}(\tilde\omega)+K_{13}(\tilde\omega)].
\end{multline}
The coefficients in Eqs.\eqref{eq:tilde_E}-\eqref{eq:h_solution} are determined at the bifurcation
point $T_b$.

Equations \eqref{eq:h}-\eqref{eq:abQ} and Eq.\eqref{eq:h_solution}
give the analytical solution for the distribution functions near
the bifurcation point. It is not difficult to find the solution of
the basic nonlinear integral equations for the distribution
functions in wider range of temperatures. This solution preserves
the shape of Eqs.(\ref{eq:h}-\ref{eq:abQ}) and
Eq.\eqref{eq:h_solution}  with temperature dependent coefficients
$a(T)$, $b(T)$ and $c(T)$ [$r(T)$ and $s(T)$]. But the relative
ratio of the molecules number [that is determined by $a(T)/b(T)$
and $r(T)/s(T)$ ratio] weakly depend on the temperature  that does
not correspond to the experimental data. The two minima in the
potential energy [correspond to $P$ and $H$ orientations] are
separated by rather high
barrier\cite{Sundqvist,Moret,Gugenberger,Yu}, that is not taken
into account in our model, Eq.\eqref{eq:D}, so we introduce
$J_P=J_P(T)$ and $J_H=J_H(T)$ to correct it.\cite{Lapinskas} It is
obvious that the pair interactions between molecules become less
sensible to their mutual orientations when the temperature
increases because the libration increases. The theoretical
estimate\cite{Yildirim} shows that the libration amplitude about
$[111]$ direction increases nearly twice when the temperature
changes from $T_g$ to $T_c$.

We tried to fit experimental data\cite{Sundqvist,Moret} to our solution,
Eqs.\eqref{eq:h}-\eqref{eq:abQ}. It followed that when $\rho_P\to 0.7$ from below then
$T_b\to150$K from above and $J_H\to 0$ [$J_H$ changes sign at this value of $\rho_P$, but
$J_P<0$]. Two types of mutual molecular orientations are profitable above this temperature: a) the
pentagon of one molecule versus the double bond of the neighbor molecule or b) the hexagon of one
molecule versus the double bond of the neighbor molecule. Below this concentration, $\rho_P=0.7$,
the frustration appears: (a)-interaction is profitable and (b)-interaction is not. Let us remind
that in our model the presence of the barriers between the energy levels is taken into account
indirectly: through the effective parameters $J_P$ and $J_H$.

Optical experiments show that the orientational disorder at low temperatures leads to the
nonhomogeneous lattice deformation, see Refs.\onlinecite{Aksenova,Aleksandrovskii} and refs.
therein. The neutron scattering data show that at low enough temperatures a relatively large
fraction of the intermolecular contacts were with double bonds pointing to pentagonal or hexagonal
faces. Only a small fraction of molecules were oriented at random. But this picture never led to
the low temperature phase, but only to partly frustrated structures.\cite{Pintschovius2} It
follows from the experimental data\cite{Sundqvist,Aksenova} that there is small anomaly at
$T\approx 150$K but it is still unclear wether this temperature corresponds to the arrest of the
free uniaxial rotation or to a glass transition similar to that observed near 90K. Note that the
discrete description of the orientational ordering, Eqs.\eqref{eq:h}-\eqref{eq:abQ}, includes as
the answer the equiprobable distribution of the molecule orientations in the cubic lattice,
$K_1(\omega)$, that corresponds to the rotations in cubic lattice [see, e.g.,
Ref.\onlinecite{Aksenov}] in addition to $H$ and $P$ states.

In summary at high temperatures the molecules travel slightly between different types of mutual
orientations. When the temperature decreases the possibility to change the orientation for the
molecule decreases because of large potential barrier between the lowest orientational states.
Then the H-state of the molecule is rather rare phenomenon at low temperatures. The glassy state
is obtained when the thermal energy is not sufficient to overcome the potential barrier that
separates the two orientational configurations \cite{Sundqvist,Moret}.The $H$-state of the
molecule [6 hexagons and 6 double bonds] is then profitable from the double bond side and is not
from the hexagon side. The $P$-state of the molecule [6 pentagons and 6 double bonds] is then
profitable from the double bond side and from the pentagon side. We see that the behavior of our
system at low temperatures is similar to the behavior of the ``dilute'' molecular crystals, e.g.
ortho-para hydrogen.\cite{Luchinskaya,Schelkacheva} $H$-states here play the role of $p-H_2$
molecules in ortho-para mixture.

The solution, Eq.\eqref{eq:h_solution} [and
Eqs.(\ref{eq:h}-\ref{eq:abQ})], correctly describes symmetrically
oriented order at all temperatures. But the coefficients $r$ and
$s$ ($a$, $b$ and $c$) should be determined from the experimental
data for $\rho_P$ and $\rho_H$. On this basis we suggest the
following model for the glass description in $\C60$.

Let us consider a system of particles on lattice sites $i, j$ with Hamiltonian
\begin{equation}
H=-\frac{1}{2}\sum_{i\neq j}J_{ij} \hat{U_i}\hat{U_j}, \label{one}
\end{equation}
where  $J_{ij}$ are quenched Gaussian interactions with zero mean,
\begin{equation} P(J_{ij})=\frac{1}{\sqrt{2\pi
J}}\exp\left[-\frac{(J_{ij})^{2}}{ 2J^{2}}\right], \label{two}
\end{equation}
with $ J=\tilde{J}/\sqrt{N}$.
\begin{gather}
\begin{split}
\hat U=&c\,[K_8^i(\tilde\omega)+K_9^i(\tilde\omega)+K_{10}^i(\tilde\omega)]
\\
+&d\,[K_{11}^i(\tilde\omega)+K_{12}^i(\tilde\omega)+K_{13}^i(\tilde\omega)],
\end{split}
\end{gather}
where $c$ and $d$ depend on the ratio of the $P$- and $H$-states [$d/c=-0.914$ when $\rho_P=0.83$,
i.e. at the pressure $P=0$, $T=T_g$].

Using replica approach we can write the free energy averaged over disorder in the form:
\begin{multline}\label{free}
\langle F\rangle_J/NkT=\lim_{n \rightarrow 0}\frac{1}{n}\max\biggl \{ \frac{t^2}{4}\sum_{\alpha}
(p^{\alpha})^{2} + \frac{t^2}{2}\sum_{\alpha>\beta} (q^{\alpha\beta})^{2}-
\\
-\ln\Tr_{\{U^{\alpha}\}}\exp\left[ \frac{t^2}{2} \sum_{\alpha}p^{\alpha}(\hat{U}^{\alpha})^2+t^2
\sum_{\alpha>\beta}q^{\alpha\beta}\hat{U}^{\alpha}\hat{U}^{\beta}\right]\biggr\}.
\end{multline}
Here $t=\tilde{J}/kT$, $\Tr(\ldots)\equiv \int_0^{2\pi}d\varphi\int_0^\pi d\cos(\theta)(\ldots)$.

The saddle point conditions for the free energy give the glass and regular order parameters
\begin{gather}\label{four}
q^{\alpha\beta}= \frac{\Tr\left[\hat{U}^{\alpha}\hat{U}^{\beta}
\exp\left(\hat{\theta}\right)\right]} {\Tr\left[\exp\left(\hat{\theta}\right)\right]} ,
\\
m^{\alpha}= \frac{\Tr\left[\hat{U}^{\alpha}\exp\left(\hat{\theta}\right)\right]}
{\Tr\left[\exp\left(\hat{\theta}\right)\right]},
\end{gather}
and the auxiliary order parameter
\begin{equation}\label{five}
p^{\alpha}= \frac{\Tr\left[(\hat{U}^{\alpha})^2 \exp\left(\hat{\theta}\right)\right]}
{\Tr\left[\exp\left(\hat{\theta}\right)\right]}.
\end{equation}
Here
\begin{equation}
\hat{\theta}=\frac{t^2}{2}
 \sum_{\alpha}p^{\alpha}(\hat{U}^{\alpha})^2+t^2
\sum_{\alpha>\beta}q^{\alpha\beta}\hat{U}^{\alpha}\hat{U}^{\beta}. \label{six}
 \end{equation}

In the replica symmetric (RS) approximation\cite{sk} the free energy (\ref{free}) has the form:
\begin{multline}
F=-NkT\left\{ t^2\frac{q^2}{4}-t^2\frac{p^2}{4}+\right.
\\
\left.\int_{-\infty}^{\infty}\frac{dz}{\sqrt{2\pi}}\exp\left(-\frac{z^2}{2}\right)\ln
\Tr\left[\exp\left(\hat{\theta}\right)\right]\right\}. \label{frs}
\end{multline}
Here
\begin{gather}
\hat{\theta}=zt\sqrt{q}\,\hat{U}+t^2\frac{p-q}{2}\hat{U}^2.
\end{gather}

The extremum conditions for the free energy ~(\ref{frs}) give the following equations for the
glass and regular order parameters:
\begin{gather}\label{qrs}
q=\int dz^G\left\{ \frac{\Tr\left[\hat{U} \exp\left(\hat{\theta}\right)\right]}
{\Tr\left[\exp\left(\hat{\theta}\right)\right]}\right\}^{2},
\\
m=\int dz^G\left\{ \frac{\Tr\left[\hat{U} \exp\left(\hat{\theta}\right)\right]}
{\Tr\left[\exp\left(\hat{\theta}\right)\right]}\right\},
\end{gather}
and the auxiliary equation
\begin{equation}
p=\int dz^G \frac{\Tr\left[\hat{U}^2 \exp\left(\hat{\theta}\right)\right]}
{\Tr\left[\exp\left(\hat{\theta}\right)\right]} \label{prs}
\end{equation}
Here
\begin{gather} \int dz^G = \int_{-\infty}^{\infty}
\frac{dz}{\sqrt{2\pi}}\exp\left(-\frac{z^2}{2}\right).
\end{gather}
\begin{figure}[thb]
\begin{center}
\includegraphics[height=80mm]{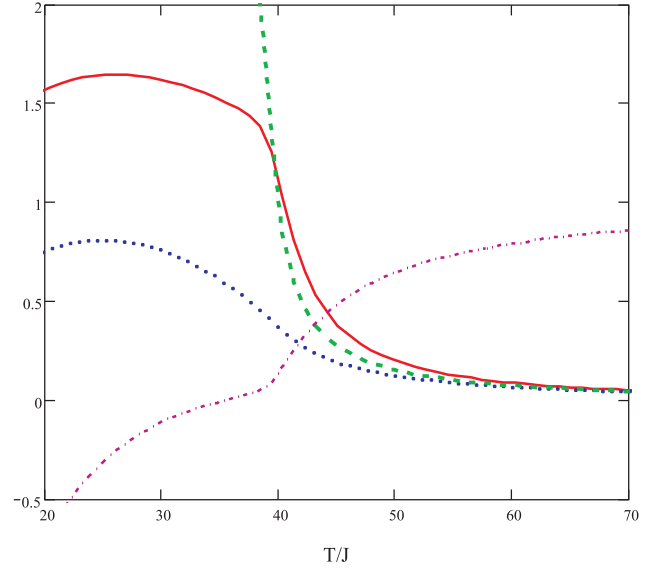} \caption{Order
parameters and the heat capacity evolution with the temperature. Here $d/c=-0.914$, red solid
curve is the heat capacity, dotted blue and dash green curves are the order parameters $m$ and
$\sqrt q$ respectively, violet dashed curve is the replicon mode $\lambda_{\rm repl}$. The replica
symmetry breaking occurs at the temperature $T_0$ corresponding to $\lambda_{\rm repl}=0$; the
glass transition temperature $T_g$ usually is very close to $T_0$.} \label{fig1}
\end{center}
\end{figure}
The replica symmetric solution is stable unless the replicon mode energy $\lambda$ is nonzero. For
our model we have:
\begin{multline}\label{lambda}
\lambda_{\rm RS}= 1 -  t^2 \times\\
\int dz^G \left\{\frac{\Tr\left[\hat{U}^2
\exp\left(\hat{\theta}\right)\right]} {\Tr\left[\exp\left(\hat{\theta}\right)\right]}-
\left[\frac{\Tr\left[\hat{U} \exp\left(\hat{\theta}\right)\right]}
{\Tr\left[\exp\left(\hat{\theta}\right)\right]}\right]^2\right\}^2.
\end{multline}

The results of the calculations are illustrated in Fig.\ref{fig1}. The order parameters do not go
to zero because $\int d\omega u^{2k+1}(\omega)\neq 0$, $k=0,1,\ldots$.\cite{4avtora} The
co-existence of the orientational ordered state (the order parameter $m$) and the glass (the order
parameter $q$) is in agrement with the experimental
data.\cite{Sundqvist,Moret,David3,David,David2}

The replica symmetry breaking occurs at the temperature $T_0$ corresponding to $\lambda_{\rm
repl}=0$; the glass transition temperature, $T_g$, usually is very close to $T_0$. The mildly
sloping curve with the broad maximum for the heat capacity qualitatively corresponds to the curve
$C_p^{\rm exp}(T)$ obtained in Ref.\onlinecite{Gugenberger}.

It is well known that the temperature of orientational transition increases with
pressure.\cite{Sundqvist,Moret} When the pressure increases the number of $P$-states decreases,
but the number of $H$-states rises. It is known that $J_P=J_H=J$ and $\rho_H=0.5$ when the
pressure $P\approx0.25\,\rm Gpa$ and $T\approx 300$K [see Fig.13 in Ref.\onlinecite{Sundqvist}].
Then we get from our model using the bifurcation condition: $J=-242$K. If we assume that $J_P$ and
$J_H$  depend linearly on $P$ then we get:
\begin{gather}\label{eq:JP}
J_P=-300(\mathrm{K})+230(\mathrm{K/Gpa})P,
\\\label{eq:JH}
J_H=-110(\mathrm{K})-530(\mathrm{K/Gpa})P.
\end{gather}
Then if $P=0.1$Gpa then we find $T_c=281$K and $\rho_P=0.56$; if $P=0.2$Gpa then we find
$T_c=293$K and $\rho_P=0.52$; if $P=0.3$Gpa then we find $T_c=309$K and $\rho_P=0.48$. These
results agree well with the experimental data\cite{Sundqvist} for $T_c$ and $\rho_P$. We believe
that our approach describes well the orientational transition at small pressure.

When $P\to 1.3$Gpa it follows from Eqs.\eqref{eq:JP}-\eqref{eq:JH} that $J_P\to 0$ [it changes its
sign here] and $J_H<0$. So our simple model for the orientational ordering becomes invalid.
Experiments show\cite{Sundqvist,Moret,David3,David,David2} that at these pressures the number of
$P$-states is very small.

The orientational glass transition is hardly experimentally seen at $P\sim0.2\,$Gpa
[$\rho_P\approx \rho_H$ at all temperatures\cite{Sundqvist}] and $P\gtrsim 1.5$Gpa [$\rho_P\ll 1$]
that does not contradict the above description of the glass transition. When $\rho_P\approx
\rho_H$ the $P$- and $H$-states are both profitable and the analogy with diluted multipole systems
becomes invalid. When the pressure is large enough $\rho_P$ becomes very small and there is no
sense speaking about disorder and frustration.

Authors thank V.N.Ryzhov for helpful discussions and valuable
comments.

This work was supported in part by the Russian Foundation for Basic Research (Grants
No.05-02-17621,05-02-17280,06-02-17519; by NWO-RFBR grant No.04-01-89005 (047.016.001), Science
Support Foundation and CRDF.

\end{document}